\documentclass[trackchanges,twocolumn]{aastex7}

\newcommand\kms{\rm km\, s^{-1}}
\usepackage[]{comment}

\usepackage{soul}
\usepackage{xcolor}

\def  \paul{}

\begin{document}

\title{Spectroscopic Confirmation: Fast rotators in the young clusters NGC~1856 and NGC~1953}

\correspondingauthor{Paul I. Cristofari}
\email{paul.cristofari@cfa.harvard.edu}

\author[0000-0003-4019-0630]{Paul I. Cristofari}
\affiliation{Center for Astrophysics $\vert$ Harvard \& Smithsonian,
	60 Garden Street, 
	Cambridge, MA 02138, United States}
\email{paul.cristofari@cfa.harvard.edu}
	
\author[0000-0002-8985-8489]{Andrea K. Dupree}
\affiliation{Center for Astrophysics $\vert$ Harvard \& Smithsonian,
	60 Garden Street,
	Cambridge, MA 02138, United States}
\email{adupree@cfa.harvard.edu}

\author[0000-0001-7506-930X]{Antonino P. Milone}
\affiliation{Dipartimento di Fisica e Astronomia ``Galileo Galilei'', Univ. di Padova, Vicolo dell’Osservatorio 3, Padova, IT-35122}
\affiliation{Istituto Nazionale di Astrofisica - Osservatorio Astronomico di Padova, Vicolo dell’Osservatorio 5, Padova, IT-35122}
\email{antonino.milone@unipd.it}

\author[0000-0002-3856-232X]{Mario Mateo}
\affiliation{Department of Astronomy, University of Michigan, Ann Arbor, MI 48109, USA}
\email{mmateo@umich.edu}

\author[0009-0006-5170-240X]{Matias Chiarpotti}
\affiliation{Instituto Interdisciplinario de Ciencias Básicas (ICB), CONICET-UNCuyo, Padre J. Contreras 1300, M5502JMA, Mendoza, Argentina}
\affiliation{Consejo Nacional de Investigaciones Cient´ıficas y Técnicas (CONICET), Godoy Cruz 2290, C1425FQB, Buenos Aires, Argentina}
\email{mati.charpo@gmail.com}

\begin{abstract}
	We present the results of a spectroscopic investigation of two Large Magellanic Cloud globular clusters, NGC~1953 and NGC~1856. Both clusters have similar ages (250 and 300\,Myr, respectively). Spectra were recorded with the Michigan/Magellan Fiber System  located on the Magellan-Clay 6.5\,m telescope. Spectra were visually inspected to assess the presence {\paul of} stellar H$\alpha$ emission lines attributed to B stars rotating close to breakup velocity (Be stars). High fractions of Be stars in the cluster typically indicate the presence of a large population of fast rotating stars, predicted by some models to explain the observed split and extended main sequence. There are numerous Be star candidates in NGC~1856, exhibiting weak but broad H$\alpha$ emission. However, only one such target was detected in NGC~1953. 
	This stark contrast between the observed populations for NGC~1856 and NGC~1953 may suggest that cluster density plays a key role in determining the fraction of Be stars. 
	These results provide essential constraints for the different scenarios attempting to explain the bimodal distribution of rotational velocities and the multiple populations of stars observed in globular clusters. 
	The impact of stellar radial velocity and nebular emission on photometric measures is assessed through simulations relying on the spectra. 
	These simulations suggest that photometric studies can under-estimate the fraction of H$\alpha$ emitters in a cluster, in particular for stars with relatively weak emission features. 
	The results also show that nebular emission has minimal impact on the  photometric  H$\alpha$ excesses.
\end{abstract}



\section{Introduction} \label{sec:intro}

Photometric studies over the last decades were marked by the discovery of extended main sequence turnoff regions (eMSTO) and split main sequence (MS) in the color magnitude diagram of globular clusters~\citep[e.g.,][]{milone-2009, milone-2013,milone-2016, correnti-2017}.
These observations provided evidence of multiple stellar populations in young clusters. Several scenarios were proposed in order to account for these observations. Some works speculated an extended star formation period in the history of the cluster formation, leading to multiple generations of stars~\citep{mackey-2008, milone-2009, conroy-2011, keller-2011, goudfrooij-2014}. The presence of massive stars rotating close to breakup velocity (Be stars) was suggested to explain the split MS observed in young clusters~\citep{bastian-2009, dantona-2017}, initially investigated by photometric studies~\citep{bastian-2017, milone-2018}, and later confirmed by spectroscopy~\citep{dupree-2017, bodensteiner-2023, cristofari-2024}. Estimating the content of Be stars in globular clusters therefore provides information on the presence of fast rotators, which is necessary to assess the role of rotation in the observation of multiple populations.

Photometric studies rely on significant excess through H$\alpha$ filters to estimate the fraction of Be stars from the inferred presence of H$\alpha$ emission. Differences in methodology can lead to differences in the reported fractions. In addition, different fractions of Be stars  have been reported for clusters of similar ages.
The present work focuses on two globular clusters located in the Large Magellanic Cloud (LMC): NGC~1856 and NGC~1953. Both clusters have similar ages, estimated to be between 250\,Myr and 300\,Myr, respectively~\citep{bastian-2017, milone-2018}, and similar mentalities~\citep[$\rm {[M/H]\sim0.4}$\,dex, ][]{milone-2023}. For NGC~1856,~\citet{bastian-2017} and~\citet{milone-2018} reported fractions of Be stars of $\sim30\%$ and $\sim20\%$ at main sequence turnoff, respectively, with~\citet{bastian-2017} estimating that their fraction likely represents a lower limit. For NGC~1953,~\citet{milone-2018} reports the possible presence of weak H$\alpha$ emitters in the cluster, {pointing out that the $m_{\rm F656N}-m_{\rm F814W}$ distribution appears wider than what would be expected by measurements errors alone.} In those studies, the identification of Be stars was made with narrow-band H$\alpha$ photometry.
The present paper reports a spectroscopic investigation of these two globular clusters to assess the extent of the  Be stellar populations.

This work relies on the analysis of high-resolution ($R>25,000$) spectra of 56 and 57 stars targeted in NGC~1856 and NGC~1953, respectively. {These stars lie near the top of the main sequence and its turn off region.} The H$\alpha$ line was observed to assess the presence of broad emission typical of the Be phenomenon. The impact of radial velocity shifts and nebular emission on photometric measurements is further investigated through simulations.
The spectroscopic data used in this work is described in Section~\ref{sec:observations}. We present the results of our analysis in Section~\ref{sec:results} and introduce our photometric simulation framework in Section~\ref{sec:simulations}. We discuss those results in Section~\ref{sec:discussion}.

\section{Observations and Data Reduction
	\label{sec:observations}}

Target selection relied on coordinates and Hubble Space Telescope (HST) photometry obtained from~\citet{milone-2018}. The properties of the observed targets are listed in Table~\ref{tab:m2fs-targets-ngc1856} and~\ref{tab:m2fs-targets-ngc1953}. Observations were obtained with the Michigan/Magellan Fiber System~\citep[M2FS;][]{mateo-2012}, {whose field of view allowed us to select targets distributed throughout the cluster}. Targets were selected {\paul among the brightest stars identified by~\citet{milone-2018}}, ensuring that no other star was found within 2'' of a given target,  unless the difference in magnitude between the two stars was greater than 1.8 in the F814W HST filter. {\paul This ensures that the spectrum of the brighter star is not impacted by that of a fainter nearby target.}

Spectra were recorded in 2019 and 2022 with the Li-H$\alpha$ filter (see Table~\ref{tab:m2fs-observations}), and reduced with our Python pipeline developed for M2FS observations~\citep{walker-2023, cristofari-2024}. This pipeline makes use of the \texttt{astropy/ccdproc} package~\citep{craig-2017} to perform bias, overscan and dark corrections on each of the 4 independent amplifiers of both CCDs. Wavelength calibration was carried out by identifying {\paul emission} lines in the ThAr spectrum obtained right before science exposures. Spectra recorded with fibers pointing to the sky were used to create a median sky spectrum. This median spectrum was used to correct for sky emission, adjusting the intensity of the sky median spectrum to remove emission features in the target spectra. This approach efficiently corrects for OH emission lines, but does not fully remove nebular emission (see Fig.~\ref{fig:ngc1856-skycorr}). {\paul Reduced spectra are made available on \texttt{Zenodo}
\footnote{\url{https://doi.org/10.5281/zenodo.16648150}}.

\begin{deluxetable*}{ccccccccc}
	\tablecaption{Targets observed in NGC~1856 with M2FS and Their Properties\label{tab:m2fs-targets-ngc1856}}
	\tablehead{\colhead{Target} & \colhead{R.A. (2000.0) [hh:mm:ss]} & \colhead{DEC (2000.0) [dd:mm:ss]} & \colhead{$m_{\rm F814W}$} & \colhead{$m_{\rm F336W}$} & \colhead{$m_{\rm F656N}$} & \colhead{Color\tablenotemark{a}} & \colhead{H$\alpha$ excess\tablenotemark{b}} & \colhead{H$\alpha$\tablenotemark{c}}}
	\startdata
	4 & 77.392031 [05:09:34] & -69.125446 [-69:07:32] & 17.3481 & 17.6337 & 17.2765 & 0.2856 & -0.0716 & e \\
	5 & 77.372056 [05:09:29] & -69.135188 [-69:08:07] & 17.8732 & 18.1901 & 17.7904 & 0.3169 & -0.0828 & e \\
	6 & 77.366109 [05:09:28] & -69.124794 [-69:07:29] & 18.0073 & 18.3546 & 17.9162 & 0.3473 & -0.0911 & e \\
	\enddata
	\tablenotetext{a}{Color defined as $m_{\rm F336W}-m_{\rm F814W}$.}
	\tablenotetext{b}{H$\alpha$ excess defined as $m_{\rm F656N}-m_{\rm F814W}$.}
	\tablenotetext{c}{Spectroscopically identified H$\alpha$ emission (e) or absorption (a).}
\tablecomments{Table~\ref{tab:m2fs-targets-ngc1856} is published in its entirety in the machine-readable format.
	A portion is shown here for guidance regarding its form and content.}
\end{deluxetable*}

\begin{deluxetable*}{ccccccccc}
	\tablecaption{Targets observed in NGC~1953 with M2FS and Their Properties\label{tab:m2fs-targets-ngc1953}}
	\tablehead{\colhead{Target} & \colhead{R.A. (2000.0) [hh:mm:ss]} & \colhead{DEC (2000.0) [dd:mm:ss]} & \colhead{$m_{\rm F814W}$} & \colhead{$m_{\rm F336W}$} & \colhead{$m_{\rm F656N}$} & \colhead{Color\tablenotemark{a}} & \colhead{H$\alpha$ excess\tablenotemark{b}} & \colhead{H$\alpha$\tablenotemark{c}}}
	\startdata
	106 & 81.310874 [05:25:15] & -68.850174 [-68:51:01] & 18.5788 & 18.8496 & 18.3124 & 0.2708 & -0.2664 & e \\
	\hline
	5 & 81.336957 [05:25:21] & -68.840059 [-68:50:24] & 17.263 & 17.9467 & 17.2778 & 0.6837 & 0.0148 & a \\
	7 & 81.357941 [05:25:26] & -68.840837 [-68:50:27] & 17.416 & 17.4142 & 17.3371 & -0.0018 & -0.0789 & a \\
	\enddata
	\tablenotetext{a}{Color defined as $m_{\rm F336W}-m_{\rm F814W}$.}
	\tablenotetext{b}{H$\alpha$ excess defined as $m_{\rm F656N}-m_{\rm F814W}$.}
	\tablenotetext{c}{Spectroscopically identified H$\alpha$ emission (e) or absorption (a).}
\tablecomments{Table~\ref{tab:m2fs-targets-ngc1953} is published in its entirety in the machine-readable format.
	A portion is shown here for guidance regarding its form and content.}
\end{deluxetable*}

\begin{deluxetable*}{ccccccccc}
	\tablecaption{List of observations\label{tab:m2fs-observations}}
	\tablehead{
		\colhead{Cluster} & \colhead{Age (Myr)} & \colhead{Date} & \colhead{Filter} & \colhead{No. Exp.} & \colhead{Exp. Time (s)} & \colhead{Total time (h)}}
	\startdata
	NGC~1953 & 250\tablenotemark{a}/350\tablenotemark{b}  & 2019 Nov 20 & Li-H$\alpha$ & 4 & 2400 & 2.7  \\
	NGC~1856 & 250\tablenotemark{a} & 2022 Nov 17 & Li-H$\alpha$ &  4/1/1 & 3600/1800/2700 &  5.25 \\
	\enddata
	\tablenotetext{a}{\citet{milone-2018}}
	\tablenotetext{b}{\citet{milone-2023}}
\end{deluxetable*}

\section{Results of spectroscopic analysis}
\label{sec:results}

{\paul Sky-corrected s}pectra were visually inspected to search for H$\alpha$ emission (see Fig.~\ref{fig:ngc1856-spectra}~and~\ref{fig:ngc1953-spectra}). Nebular emission is observed in both clusters, confirmed by its presence in spectra recorded with fibers pointing at the sky. Our sky correction routine removes an average nebular feature, but residuals remain in the sky-corrected spectra. The presence of stellar H$\alpha$ emission is assessed by searching for broad lines, that are wider than the typical nebular emission.

\subsection{NGC~1856}
For NGC~1856, three targets exhibit strong and broad H$\alpha$ emission features (29, 100, 102, see Fig.~\ref{fig:ngc1856-spectra}). These stars also show clear excess in the photometric data (Fig.~\ref{fig:ngc1856-halpha-photo}). The position of the 3 targets fall to the blue of the main sequence in the CMD (see Fig.~\ref{fig:ngc1856-photo}). Such blue populations of stars, whose position are compatible with young field stars, have been observed in other clusters such as NGC~1783, where they were proposed to be either younger stars originating from an extended formation period within the cluster, or to be blue stragglers belonging to the cluster~\citep{milone-2023}.
 Ten other targets show weaker but clear emission, typically in the middle of a broader absorption line. 

H$\alpha$ candidates are identified from photometry following~\citet{milone-2018}. A fiducial line is defined {\paul in the color axis $m_{\rm F656N}-m_{\rm F814W}$} representing the centroid of the stars with no H$\alpha$ emission, and targets deviating by more than $-0.15$ from this fiducial line are considered candidates with strong H$\alpha$ emission (see Fig.~\ref{fig:ngc1856-halpha-photo}). Out of the 13 targets with clear H$\alpha$ emission in their spectra, 6 are found closer to the fiducial line than $-0.15$, thus suggesting that~\citet{milone-2018} technique underestimates the presence of H$\alpha$ emitters.

\begin{figure}
	\includegraphics[width=.95\linewidth]{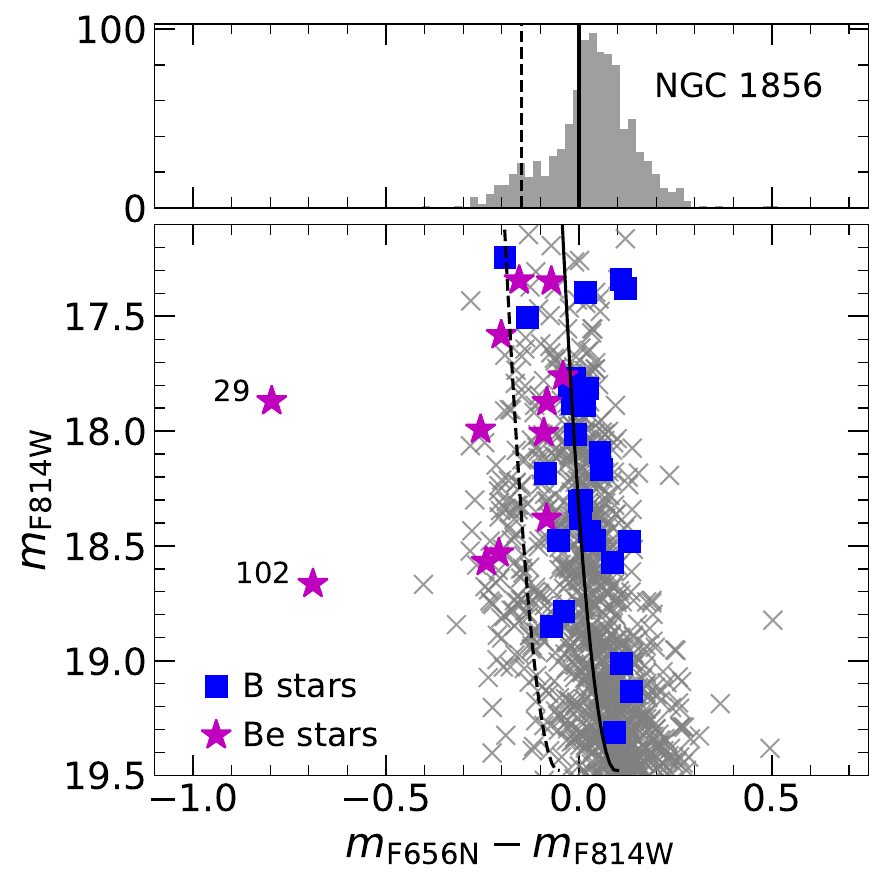}
	\caption{Photometric H$\alpha$ excess for NGC~1856. Gray X symbols present the photometric data from~\citet{milone-2018} for stars located less than 20 arcsec from the cluster center. Following~\citet{milone-2018}, only stars with $-0.25<m_{\rm F336W}-m_{\rm F814W}<1.5$ are considered in order to avoid young field objects.  Magenta stars mark the position of H$\alpha$ emitters identified through spectroscopy, and blue squares show the position of the other observed targets. Following~\citet{milone-2018}, a fiducial line (solid line) is drawn by eye to indicate the central position of the stars with no H$\alpha$ emission. The dashed line indicates the $-0.15$ threshold relative to this line, used to identify H$\alpha$ emitter candidates. The top panel shows a histogram representing the photometric data by~\citep{milone-2018}. {\paul Targets 29 and 102 are identified on the figure; no HST H$\alpha$ excess was available for target 100 in spite of the strong stellar H$\alpha$ emission line observed through spectroscopy (see full Fig.~\ref{fig:ngc1856-spectra} available online).}}
	\label{fig:ngc1856-halpha-photo}
\end{figure}

\begin{figure}
	\includegraphics[width=.95\linewidth]{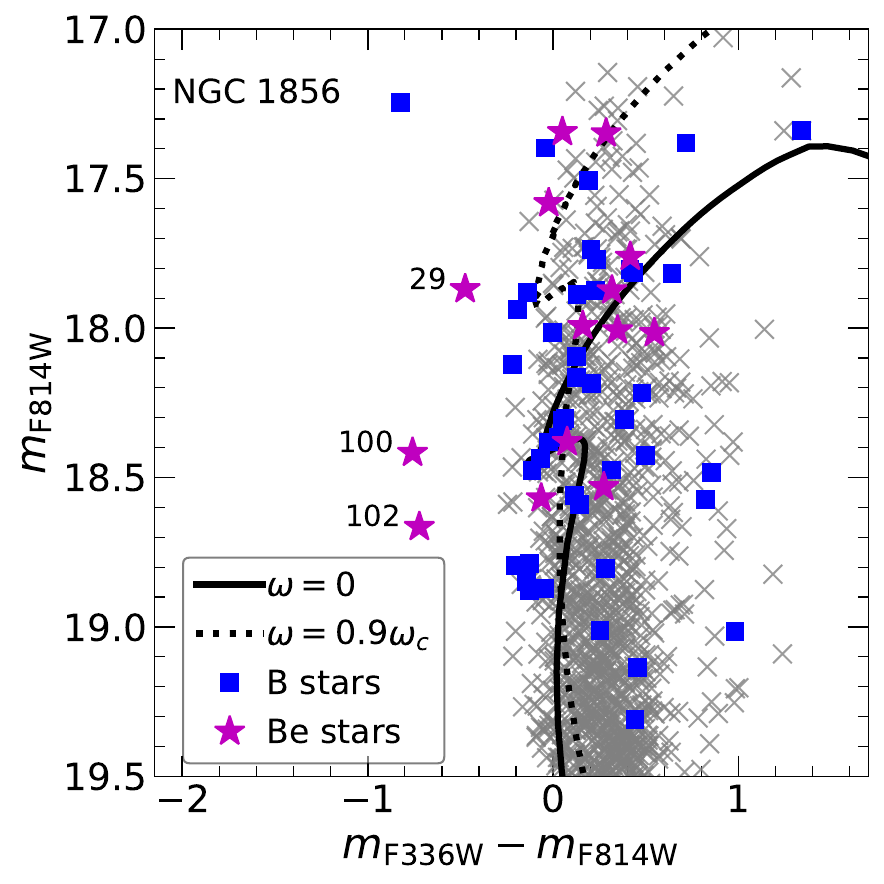}
	\caption{Color magnitude diagram for NGC~1856. Photometry from~\citet{milone-2018} is indicated by gray X symbols  for stars located less than 20 arcsec from the cluster center.  Magenta stars mark the position of H$\alpha$ emitters identified through spectroscopy, and blue squares show the position of the other observed targets.
		The solid and dashed lines mark the isochrones computed by~\citet{dantona-2015} for non-rotating and fast-rotating (90\% of their critical velocity) stars, respectively, assuming an age of $\sim320$\,Myr.  {\paul Targets 29, 100 and 102 are identified on the figure.}}
	\label{fig:ngc1856-photo}
\end{figure}

The photometric measurements for these stars are compared to evolutionary tracks computed by~\citet{dantona-2015} for non-rotating and fast-rotating (90\% of their critical velocity) stars assuming an age of $\sim$~$320$\,Myr (see Fig.~\ref{fig:ngc1856-photo}).
These models predict a higher fraction of fast rotators at bright magnitudes assuming a single age for all stars. One should note that a younger non-rotating population can also mimic an older population of fast-rotators~\citep{dupree-2017}. 
For NGC~1856, dividing our sample in faint and bright targets at $m_{\rm F814W}=18$, we find the Be fraction of the total stars to be $Be/(B+Be)\approx0.19\pm0.03$ and $Be/(B+Be)\approx0.26\pm0.05$ for  $18<m_{\rm F814W}<19$ and $17<m_{\rm F814W}<18$, respectively.
This higher fraction of Be stars at brighter magnitudes is consistent with the fractions typically reported for other globular clusters and NGC~1856~\citep{milone-2018}, although our small sample yields relatively large error bars. 

The ratio of Be stars reported by~\citet{bastian-2017} is of $\sim30\%$ while~\citet{milone-2018} reported a ratio of  $\sim 20$\% for $m_{\rm F814W}<19$.~\citet{bastian-2017} indicates that the reported ratio likely represents a lower limit. While~\citet{bastian-2017} and~\citet{milone-2018} rely on different combinations of filters to define a continuum ($\frac{m_{\rm F555W}+m_{\rm F814W}}{2}$ and $m_{\rm F814W}$, respectively), the main differences in methodology rely on the convention used to identify H$\alpha$ emitters.~\citet{milone-2018} traces a fiducial line through the non-H$\alpha$ emitters, and searches for targets falling more than $-0.15$ away from this line.~\citet{bastian-2017} relied on the estimated error bars to identify targets that are at least 3$\sigma$ away from a fiducial line (see Fig.~\ref{fig:bastian-photo}). If the estimated errors are small, the second approach may lead to slightly larger counts of H$\alpha$ emitters.

It should be noted that spectroscopic studies such as~\citet{bastian-2017} and~\citet{milone-2018} typically focus on the core of the cluster (22 and 44 arcsec, respectively, see Fig.~\ref{fig:ngc1856-pos}). This study includes targets that extend up to 90 arcsec from the cluster center, and generally avoids the crowded cluster core.

\begin{figure}
	\includegraphics[width=\linewidth]{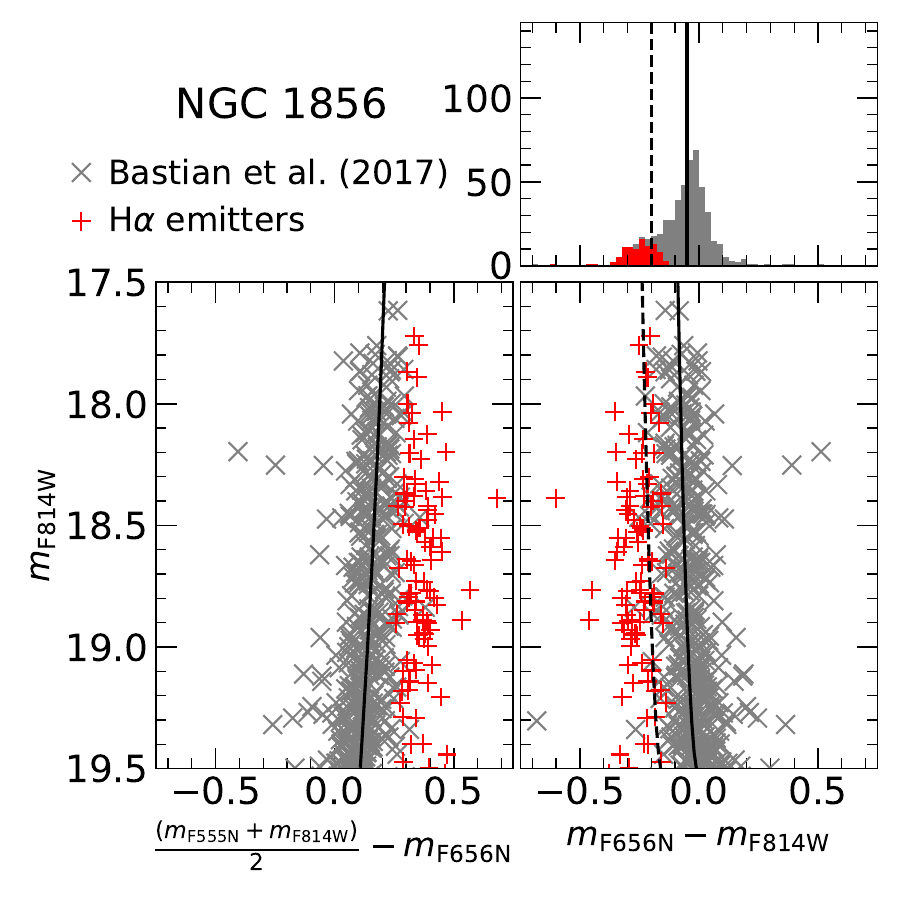}
	\caption{Photometry extracted from~\citet{bastian-2017} and presented with different combinations of photometric colors. The left panel presents data with the H$\alpha$ emitters (red crosses) identified by~\citet{bastian-2017}. The bottom right panel presents the same data with the $m_{\rm F656N}-m_{\rm F814W}$ colors typically used in~\citet{milone-2018}, with the $-0.15$ limit indicated by a dashed line. The top right panel shows a histogram obtained with the data shown in the bottom right panel. Both techniques appear to yield similar assessment of the presence of H$\alpha$ emission.}
	\label{fig:bastian-photo}
\end{figure}

\begin{figure}
	\includegraphics[width=\linewidth]{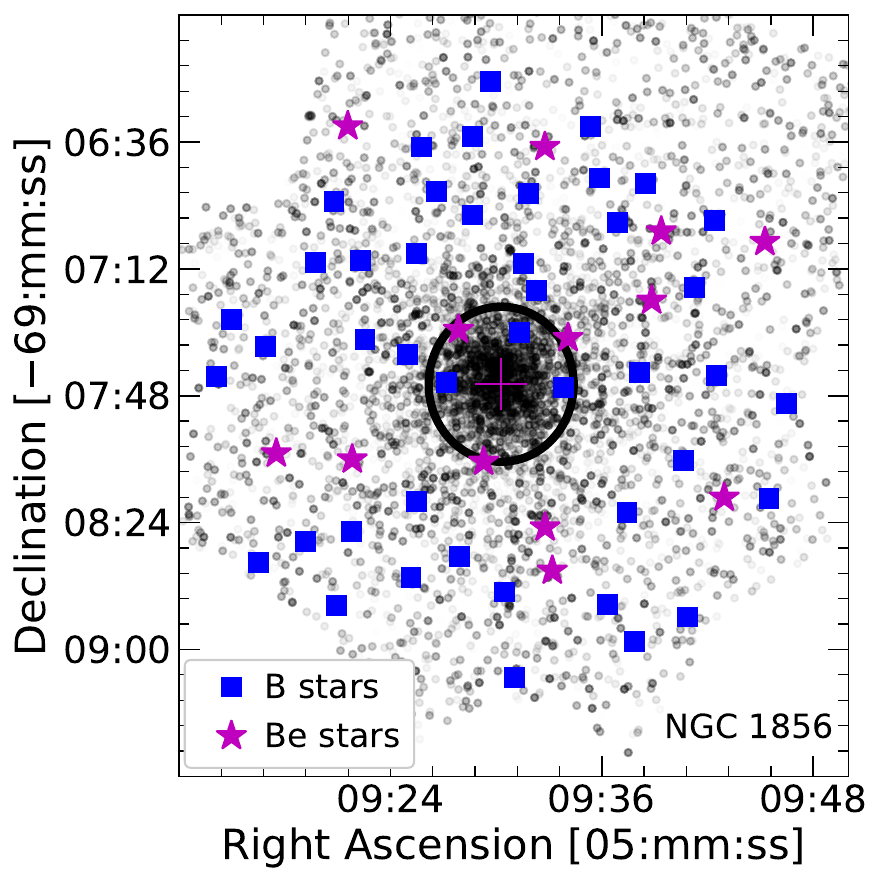}
	\caption{Position of the targets observed by~\citet{milone-2018} with $16<m_{\rm F814W}<20$. Darker colors correspond to brighter magnitudes. Magenta stars mark the position of H$\alpha$ emitters identified through spectroscopy, and blue squares show the position of the remaining observed targets. The solid circle marks the 22 arcsec circle from the cluster center (magenta cross) used by~\citet{bastian-2017} to observe only the core of the cluster.~\citet{milone-2018} estimated the content of Be stars in a similar 20 arcsec radius around the cluster center.}
	\label{fig:ngc1856-pos}
\end{figure}

\subsection{NGC~1953}

~\citet{milone-2018} did not report fractions of H$\alpha$ emitter candidates for NGC~1953, but noted that the $m_{\rm F656N} - m_{\rm F814W}$ color distribution was wider than accounted for by the measurement errors, suggesting the presence of weak H$\alpha$ emitters.

Out of the 57 stars observed, only one target was identified with weak but broad H$\alpha$ emission features in NGC~1953 ({\paul see target 61 on} Fig.~\ref{fig:ngc1953-spectra}). This target is found in the bulk of the main sequence of  the CMD (see Fig.~\ref{fig:ngc1953-photo}). The analysis of the stellar spectra is however limited by the low signal-to-noise ratio of the observations, due to fainter magnitudes.

\begin{figure}
	\includegraphics[width=.95\linewidth]{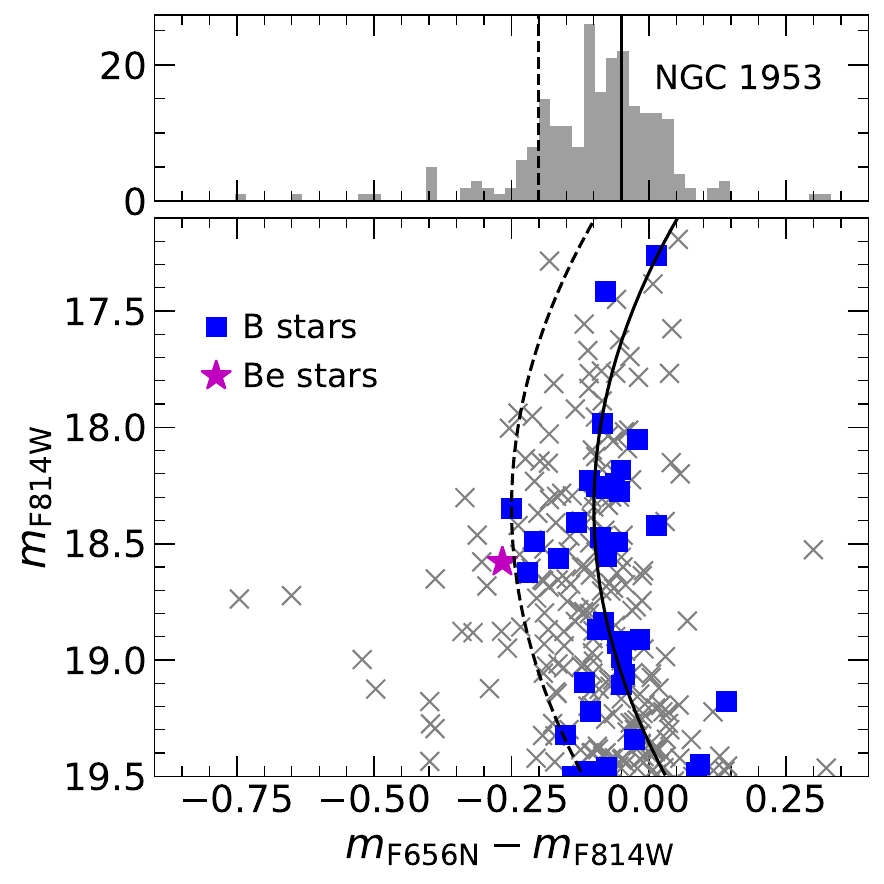}
	\caption{Photometric H$\alpha$ excess for NGC~1953. Gray X symbols present the photometric data from~\citet{milone-2018} for stars located less than 20 arcsec from the cluster center. 
		Magenta stars mark the position of H$\alpha$ emitters identified through spectroscopy, and blue squares show the position of the other observed targets. Following~\citet{milone-2018}, a fiducial line (solid line) is drawn by eye to indicate the central position of the stars with no H$\alpha$ emission. The dashed line indicates the $-0.15$ threshold about this line, used to identify H$\alpha$ emitter candidates. The top panel shows a histogram representing the photometric data~\citep{milone-2018}, with the solid and dashed vertical lines showing the position of the median of the solid and dashed lines of the lower panel along the $m_{\rm F814W}$ axis.}
	\label{fig:ngc1953-halpha-photo}
\end{figure}

\begin{figure}
	\includegraphics[width=.95\linewidth]{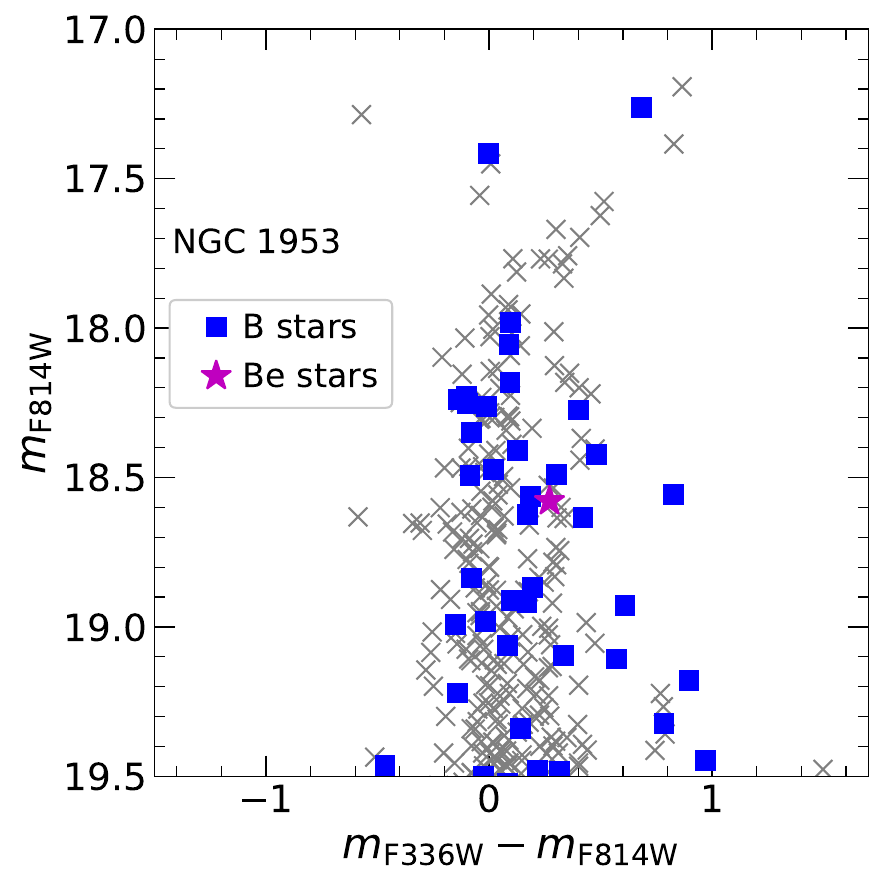}
	\caption{Color magnitude diagram for NGC~1953. Photometry from~\citet{milone-2018} is indicated by gray X symbols.   Magenta stars mark the position of H$\alpha$ emitters identified through spectroscopy, and blue squares show the position of the other observed targets without H$\alpha$ emission.}
	\label{fig:ngc1953-photo}
\end{figure}

\section{Synthetic photometry}
\label{sec:simulations}
In both NGC~1856 and NGC~1953, nebular emission is clearly observed in the spectra recorded with the fibers pointing at the sky. In addition, the LMC has a radial velocity of $\sim+260$~$\kms$~\citep{mcconnachie-2014}, placing the H$\alpha$ line towards the edge of the HST F656N filter (see Fig.~\ref{fig:example-shift-filter}). Both these phenomena can have an impact on photometric measurements. In this section, we assess the impact of nebular emission and radial velocity by estimating the excess expected through the F656N filter from normalized spectra.

\begin{figure}
	\includegraphics[width=\linewidth]{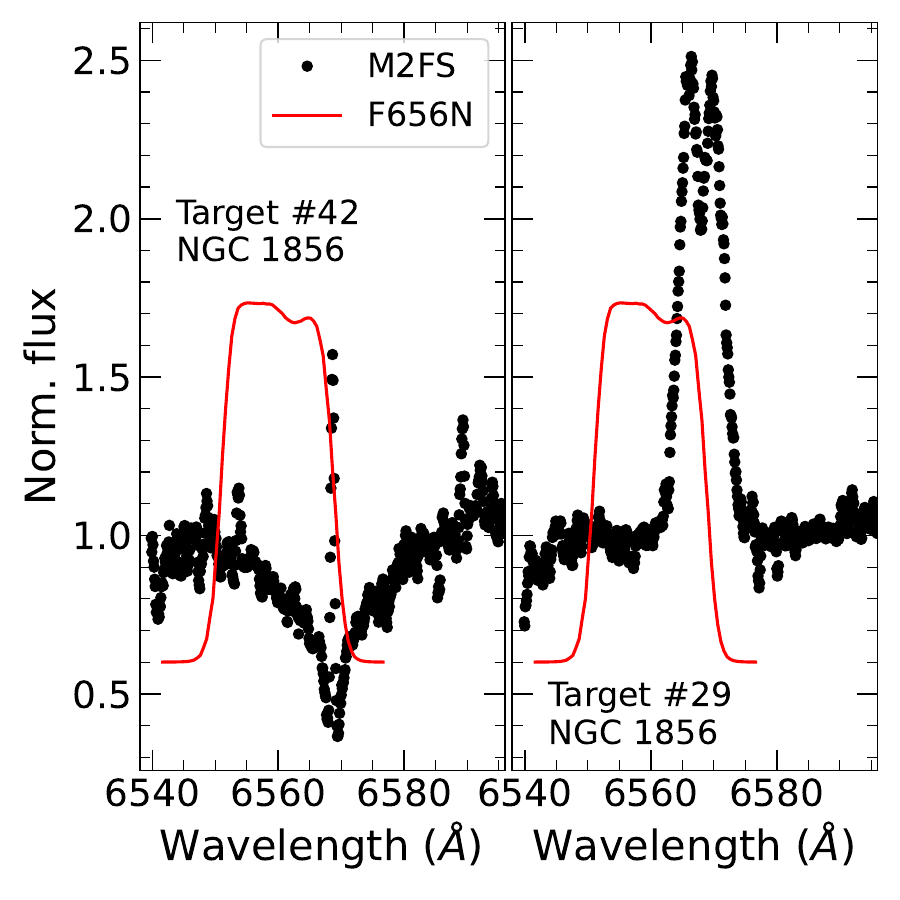}
	\caption{Comparison between the HST F656N filter response and M2FS spectra. The filter's response was re-normalized and offset for easier visual comparison. The filter was converted to air wavelengths. The high velocity shift of NGC~1856 ($\sim +260~\kms$) places the H$\alpha$ line close to the filter edge.}
	\label{fig:example-shift-filter}
\end{figure}

\subsection{Simulating the photometry excess through the HST F656N filter}

In both \citet{milone-2018} and \citet{bastian-2017}, cluster photometry to detect H$\alpha$ emission relies on the use of the F656N filter on the HST Wide Field Camera 3 (WFC3). This narrow band filter (passband rectangular width of 17.6\,\r{A}\footnote{https://hst-docs.stsci.edu/wfc3ihb/chapter-6-uvis-imaging-with-wfc3/6-5-uvis-spectral-elements}), when used on high  velocity targets, may compromise the interpretation of the measurements~\citep{dupree-2020}.

The simulations rely on the transmission curve of the F656N filter obtained from calibrations\footnote{Transmission curves can be retrieved at \url{https://www.stsci.edu/hst/instrumentation/wfc3/performance/throughputs}, and are also available on the Spanish Virtual Observatory \url{http://svo2.cab.inta-csic.es/svo/theory/fps3/index.php?id=HST/WFC3_UVIS1.F656N}}.
Excess through the HST F656N filter is estimated by computing the integral of the normalized flux through the response of the filter. The reference continuum excess is computed by performing the same integral on a normalized spectrum with no absorption lines (i.e., a continuum of 1).
To mitigate the impact of normalization on the photometry computed from spectra, we adjust the continuum with a line passing through two points defined on each side of the H$\alpha$ line. This approach reduces slopes in the continuum that could impact the results.

\subsubsection{Impact of Radial velocity}

\begin{figure*}
	\includegraphics[width=\linewidth]{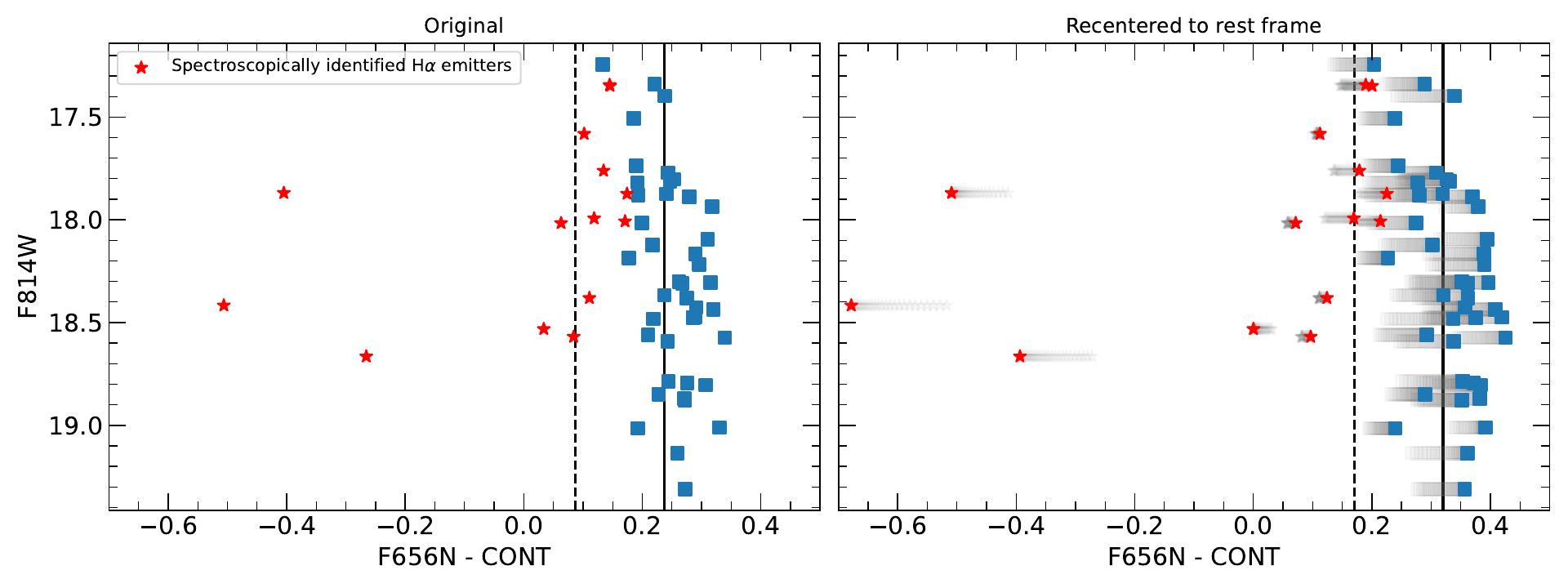}
	\caption{Impact of radial velocity on simulated excess through the F656N filter, relying on the M2FS spectra recorded for NGC~1856. The left and right panels present the computed excess through the filter before and after recentering the spectra in the rest frame, respectively. The excess through F656N filter response is plotted against the photometry obtained by~\citet{milone-2018} in the F814W filter (y axis). Red stars mark the position of the targets that display non-nebular H$\alpha$ emission features in the M2FS spectra. {\paul Blue squares indicate the stars with no identified stellar H$\alpha$ emission. While some points overlap, the total number of points (blue and red) remains 56 in both panels. The gray trail on the right panel connect the points to their initial position before recentering the spectra in rest frame.} The black line marks the median of the sample, and the dashed line indicates the fiducial boundary at $-0.15$ from the median. These simulations suggest photometric studies relying on the F656N filter tend to underestimate the number of H$\alpha$ emitters. In this simulation, the underestimate amounts to more than 50\%.}
	\label{fig:simu-rv-halpha}
\end{figure*}

To estimate the impact of radial velocity on the photometric measurements two sets of measurements are performed. First the synthetic photometric excess through the HST F656N filter is obtained for the {\paul 56} normalized spectra recorded for NGC~1856.  A second set of excess measurements is obtained on the same data after re-centering the spectra in the rest frame (by applying a shift of $+260$\,$\kms$). The obtained synthetic photometric excesses are presented in Fig.~\ref{fig:simu-rv-halpha}.

The synthetic photometric measurements derived here, while not directly comparable to observational data, are consistent with our identification of H$\alpha$ emitters. 

A consequence of the recentering the spectra is an immediate shift of most targets towards lower H$\alpha$ excesses. This can be interpreted as the consequence of absorption lines moving to the center of the filter. Consequently, the median of the sample shifts from 0.24 to 0.32. A second consequence is the shift of targets with high H$\alpha$ excess towards larger excess, resulting from broad emission features being reentered in the filter. Finally, for a number of targets with broad H$\alpha$ emission features centered in broad H$\alpha$ absorption lines, the excess remains relatively constant before and after correcting for radial velocity.
After correction, the standard deviation of the colors of the sample increases to 0.21 against 0.15 before correction.
Observations secured at different epochs can also be expected to lead to variations in the observed excesses due to the Earth (up to $\pm 30$\,$\kms$) and HST (up to $\pm 7.5$\,$\kms$) motion.

\begin{table}
	\caption{Synthetic photometric excess computed}
	{\center
	\begin{tabular}{cccc}
		\hline
		\hline
				Target & $m_{\rm F814W}$ & $(m_{\rm F656N}-C)_{\rm 0}$ & $(m_{\rm F656N}-C)_{\rm corr}$ \\
		\hline
	   3 &  17.3400 &   0.2208 &   0.2892  \\ 
		4 &  17.3481 &   0.1456 &   0.2005 \\ 
		5 &  17.8732 &   0.1742 &   0.2254 \\ 
		\hline
	\end{tabular}}
	\label{tab:simu}
	\tablecomments{{\paul The third and fourth column list the relative photometric excesses computed before and after recentering the spectra to the rest frame, respectively (see Sec.~\ref{sec:simulations}). Table~\ref{tab:simu} is published in its entirety in machine-readable format.}}
\end{table}

\subsubsection{Impact of nebular emission}

\begin{figure*}
	\includegraphics[width=\linewidth]{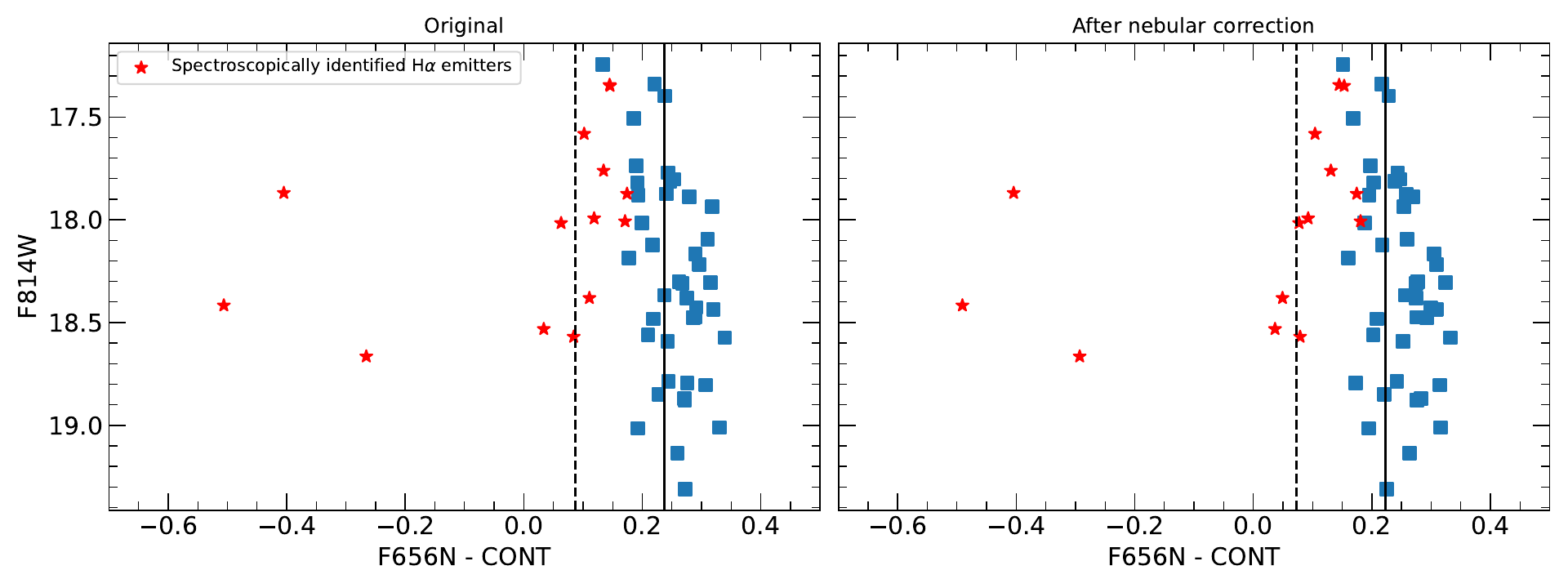}
	\caption{{\paul Same as Fig.~\ref{fig:simu-rv-halpha} exploring the impact of nebular emission.} This simulation suggests that nebular emission does not have a significant effect on the determination of the presence of stellar H$\alpha$ emission.}
	\label{fig:simu-nebu-halpha}
\end{figure*}

The impact of nebular emission on photometric measurements is tested in a similar way. A first set of measurements is obtained from the spectra for NGC~1856 before any correction of the nebular emission features. A correction of the nebular emission is performed by rejecting the bins found in the expected nebular region, and replacing these values with a linear interpolation of the spectra on this short wavelength range. It should be noted that this method is not capable of performing optimal correction of the nebular features, nor does it disentangle the stellar and nebular emission. Instead the results of this simulation provide insights on the impact of narrow features on photometry.

The synthetic photometry computed after removing the nebular features remains consistent with that obtained before correction (see Fig.~\ref{fig:simu-nebu-halpha}). These results suggest that the impact of nebular emission remains low on the recorded photometry due to the narrow nature of the features.  

\section{Discussion and conclusions}
\label{sec:discussion}

This work presents a spectroscopic investigation of the content of Be stars of the NGC~1856 and NGC~1953 Large Magellanic Cloud globular clusters. 
NGC~1856 and NGC~1953 are located in stronger H$\alpha$ emission regions than clusters such as NGC~1866 (see Fig.~\ref{fig:lmc-halpha}). These features render characterization and detection of stellar H$\alpha$ emission challenging. Spectra were recorded with M2FS for both clusters, and nebular emission was confirmed through spectra recorded with fibers pointing at the sky.

Spectra recorded with M2FS were visually inspected to assess the presence of stellar H$\alpha$ emission, attributed to the presence of B stars rotating close to breakup velocity (Be stars).  The recorded spectra reveal more than ten targets with clear H$\alpha$ emission features, confirming a high fraction of Be stars in this cluster. While photometric studies reported fractions of $\sim 30\%$ or $\sim 20\%$ for NGC~1856~\citep{bastian-2017, milone-2018}, several H$\alpha$ emitters confirmed through spectroscopy show weak emission through the HST F656N filter and were therefore not classified as H$\alpha$ emitters with photometry. These results support those of~\citet{bastian-2017} suggesting the true fraction of Be stars may be larger, as some targets with decretion disks may not lead to detectable H$\alpha$ emission due to insufficient ionizing flux.

Only one clear H$\alpha$ emitter was observed in NGC~1953, although spectroscopic analysis of the cluster is hampered by the limited signal-to-noise ratio of the observations.
Those results appear to suggest a significant difference in the fraction of Be stars in NGC~1953 compared to NGC~1856, despite the similar age of the two clusters. 
This puzzling result may highlight the role of cluster density, with the mass of NGC~1953 reported to be $\sim$13\% that of NGC~1856~\citep{milone-2018}. The idea that higher cluster densities at the time of star formation could favor higher Be star contents has been proposed~\citep{strom-2005, wolff-2007}.
	 Our results directly support recent findings, including those of~\citet{bastian-2017} who concluded from the presence of fast rotators in older clusters than those previously studied~\citep{huang-2010} that density plays a key role in determining the presence of a population of fast rotators at different ages. 
	 Recent works suggested that binary interaction can explain the CMD of young globular clusters~\citep{demink-2013, hastings-2021, bodensteiner-2023}, in particular through mass-transfer. Stellar density could play a key role in both scenarios, impacting either the early stages of cluster formation, or binary interactions.
	 {\paul The observation of Be stars whose position in the CMD are distributed throughout the eMSTO and the presence of potential blue stragglers exhibiting strong H$\alpha$ emission support the role of binary interactions in the observed populations. The recent spectroscopic investigations of young SMC and LMC clusters~\citep[e.g.,][]{dupree-2017, bodensteiner-2023, kamann-2023, cristofari-2024} provide strong observational support a rotation-driven scenario to explain the shape of the eMSTO, with now clear independent indications of bimodal distribution in the stellar rotational velocities in several young clusters. One should note that the case of NGC~330 stands out~\citep{bodensteiner-2020,bodensteiner-2021,bodensteiner-2023,  cristofari-2024}, with a particularly visible split sequence near the eMSTO, largely populated by Be stars. In contrast to other LMC clusters, NGC~330 is both very young ($\sim 40$\,Myr) and formed in a low-metallicity environment. To assess more thoroughly the role of metallicity and age on stellar populations, further spectroscopic investigations of SMC clusters are needed.}

{\paul Our results suggest that denser} clusters would favor the formation of Be stars. If cluster density is indeed a key parameter in determining the fraction of Be stars, spectroscopic observations of other clusters with properties similar to those of NGC~1953, such as NGC~1801, should reveal very few H$\alpha$ emitters. Such observations would provide important constraints for models attempting to explain the observations of multiple populations.

The impact of radial velocity and nebular emission on photometric H$\alpha$ excess was investigated through simulations relying on the recorded M2FS spectra. Radial velocity shifts have a significant impact on the photometric, and re-centering spectra to the rest frame led to a 40\% increase of the dispersion in our synthetic photometry. These results further support that photometric studies are likely to underestimate the ratio of Be stars in clusters such as NGC~1856, as these rely on the dispersion of the data and distance between H$\alpha$ and non-H$\alpha$ emitters in the CMD.
In contrast, the nebular emission appears to have limited impact on the photometric results, as emission features are sufficiently narrow, and observed in most spectra.
Future photometric observations aimed at studying H$\alpha$ emission in LMC globular clusters could adapt their strategies by relying on different filters, such as the broader F657N HST filter. {\paul While the use of a broader filter may result in larger uncertainties on H$\alpha$ estimates, strategies combining F656N and F657N observations, along with simulation of the expected excess through both filters, would allow one to devise the best observing strategy for LMC clusters.}

Future spectroscopic observations with high-resolution instruments should be able to confirm the high ratio of Be stars in NGC~1856, and enable the detailed characterization of objects with H$\alpha$ emission lines. These are crucial to establish a well constrained set of parameters for a large sample of stars, and assess the role of rotation in the birth of a of a large population of fast rotating B stars within young massive clusters.

\begin{acknowledgments}
	This research has made use of ``Aladin sky atlas'' developed at CDS, Strasbourg Observatory, France~\citep{aladin-a, aladin-b, aladin-c}.
	
	This work has been funded by the European Union~–~NextGenerationEU RRF M4C2 1.1 (PRIN 2022 2022MMEB9W: “Understanding the formation of
	globular clusters with their multiple stellar generations”, CUP C53D23001200006, PI Anna F. Marino).

	We thank Dr. Jennifer Yee for her help in extracting right ascension and declination from HST images for the data published by~\citet{bastian-2017}.
	
	We thank Dr. Emily Leiner and Dr. Morgan Macleod for fruitful discussions regarding the results of this paper.
\end{acknowledgments}

\begin{contribution}

AKD proposed the observations. APM provided photometric data. MM took the observations. MC contributed to the reduction of the spectra. PIC performed the spectroscopic analysis and wrote the manuscript. All authors contributed to the interpretation of the results.


\end{contribution}

%
\facility{Magellan: Clay (M2FS, MIKE).}

\software{
      astropy / ccdproc \citep{astropy-2013, craig-2017},
	 Numba~\citep{lam-2015}.
}

\appendix

\section{H$\alpha$ emission}
Figure~\ref{fig:lmc-halpha} presents an image showing nebular H$\alpha$ emission in the direction of the LMC.

\begin{figure*}
	\includegraphics[width=0.99\linewidth]{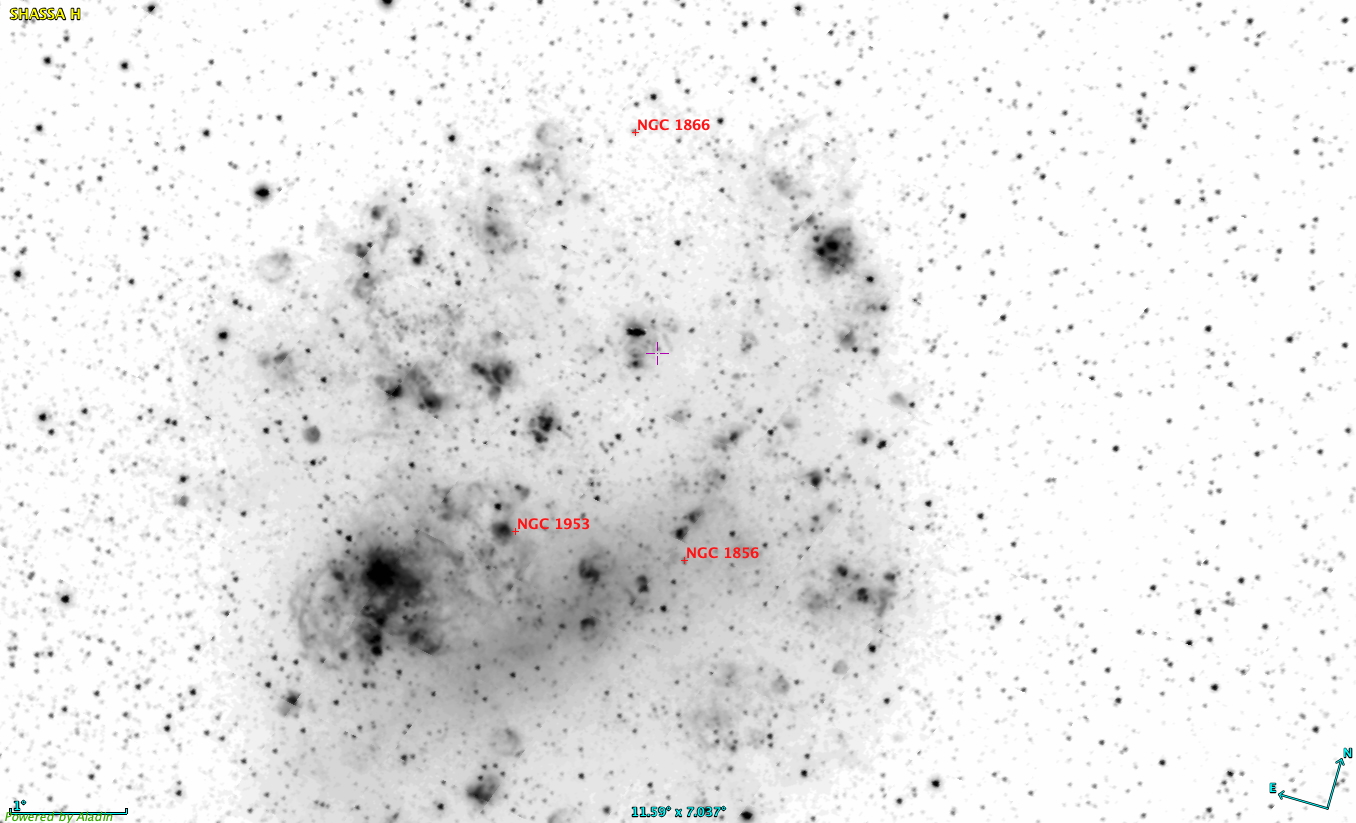}
	\caption{NGC~1866, NGC~1953 and NGC~1856 seen in the LMC. The Southern H-Alpha Sky Survey Atlas (SHASSA) image reveals nebular emission throughout the LMC, and in particular in the regions where NGC~1953 and NGC~1856 are found.}
	\label{fig:lmc-halpha}
\end{figure*}

\section{M2FS spectra}
Figures~\ref{fig:ngc1856-spectra} and~\ref{fig:ngc1953-spectra} present the spectra obtained with M2FS for NGC~1856 and NGC~1953. {\paul Figures~\ref{fig:ngc1856-skycorr} shows an example of sky correction for NGC~1856.}

\figsetstart
\figsetnum{11}
\figsettitle{Spectra recorded with M2FS for NGC~1856.}

\figsetgrpstart
\figsetgrpstart
\figsetgrpnum{11.1}
\figsetgrptitle{Target 3 to 33}
\figsetplot{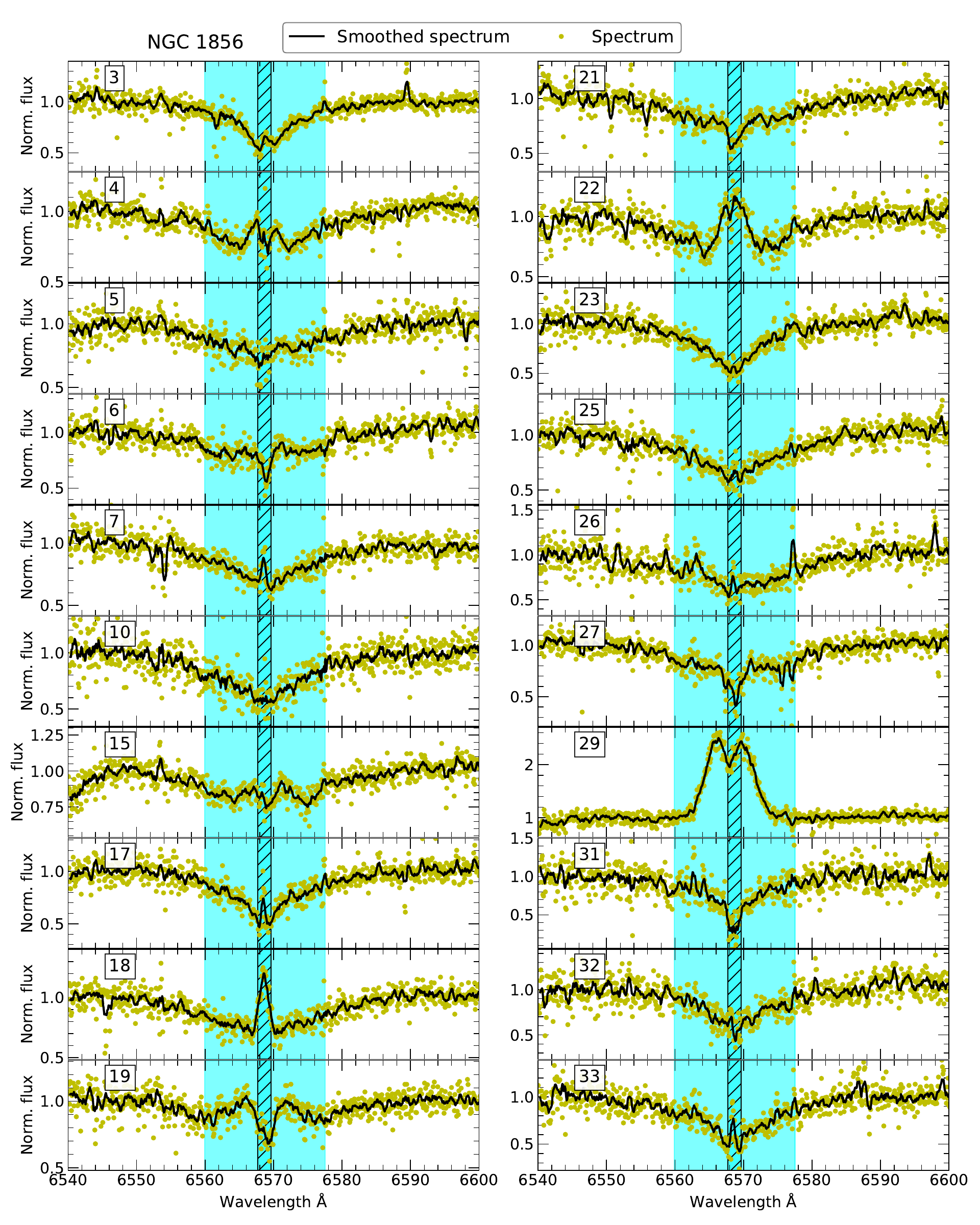}
\figsetgrpnote{}
\figsetgrpend

\figsetgrpstart
\figsetgrpstart
\figsetgrpnum{11.2}
\figsetgrptitle{Target 34 to 66}
\figsetplot{figures/ngc1856_plots_2.pdf}
\figsetgrpnote{}
\figsetgrpend

\figsetgrpstart
\figsetgrpstart
\figsetgrpnum{11.3}
\figsetgrptitle{Target 67 to 102}
\figsetplot{figures/ngc1856_plots_3.pdf}
\figsetgrpnote{}
\figsetgrpend

\figsetend

\begin{figure*}
	\centering
	\includegraphics[width=.99\linewidth]{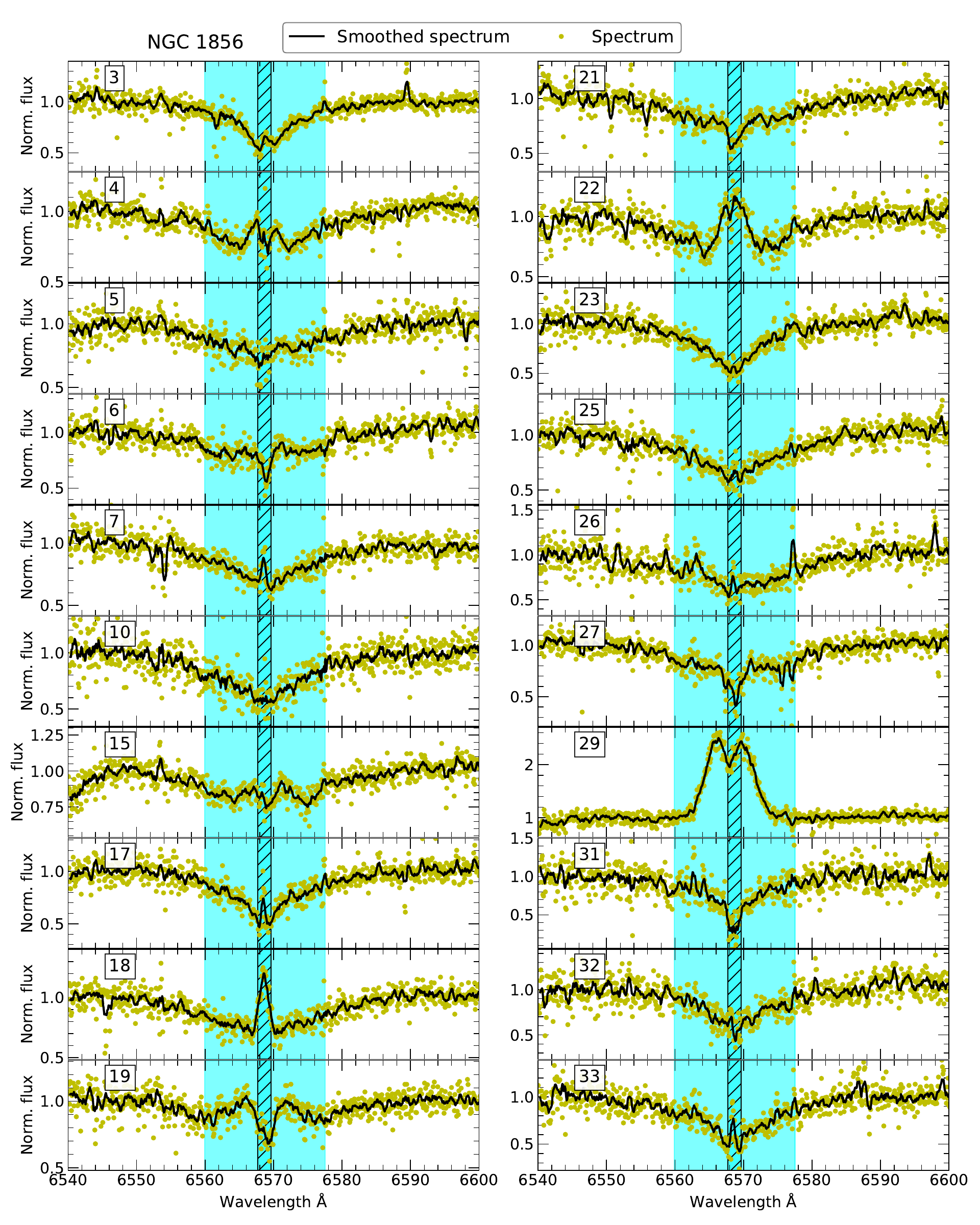}
	\caption{Spectra recorded with M2FS for NGC~1856. {\paul The cyan regions indicates width of the HST H$\alpha$ F656N filter}. The {\paul hatched} regions indicates the typical width of the nebular emission recorded in the sky spectra. {\paul The yellow points and black line show the spectrum before and after smoothing, respectively. The target number is specified in each panel.} The complete figure set (3 figures) is available in the online journal.}
	\label{fig:ngc1856-spectra}
\end{figure*}

\figsetstart
\figsetnum{12}
\figsettitle{Spectra recorded with M2FS for NGC~1953.}

\figsetgrpstart
\figsetgrpstart
\figsetgrpnum{12.1}
\figsetgrptitle{Target 5 to 117}
\figsetplot{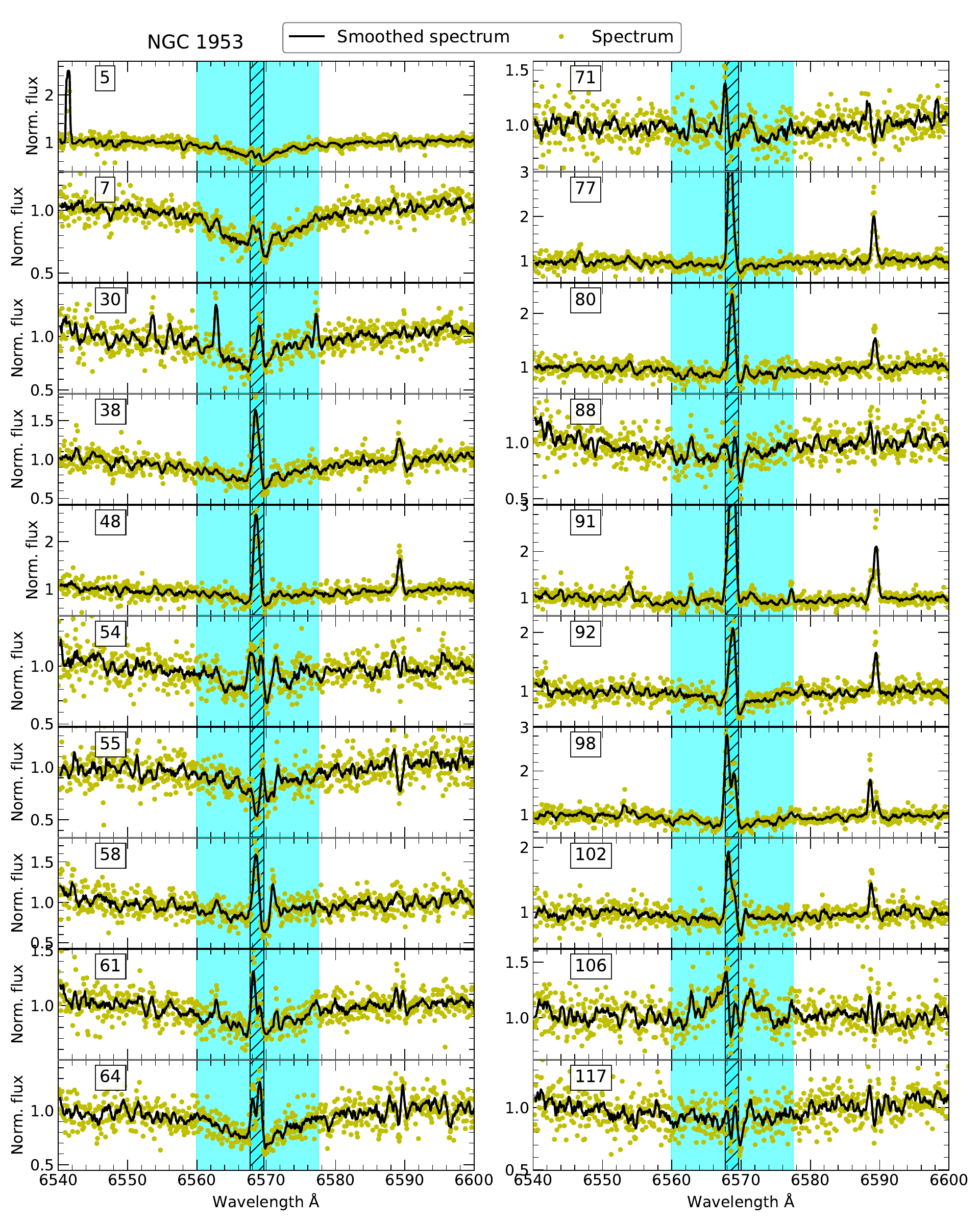}
\figsetgrpnote{}
\figsetgrpend

\figsetgrpstart
\figsetgrpstart
\figsetgrpnum{12.2}
\figsetgrptitle{Target 120 to 374}
\figsetplot{figures/ngc1953_plots_2.pdf}
\figsetgrpnote{}
\figsetgrpend

\figsetgrpstart
\figsetgrpstart
\figsetgrpnum{12.3}
\figsetgrptitle{Target 383 to 693}
\figsetplot{figures/ngc1953_plots_3.pdf}
\figsetgrpnote{}
\figsetgrpend

\figsetend

\begin{figure}
	\centering
	\includegraphics[width=.99\linewidth]{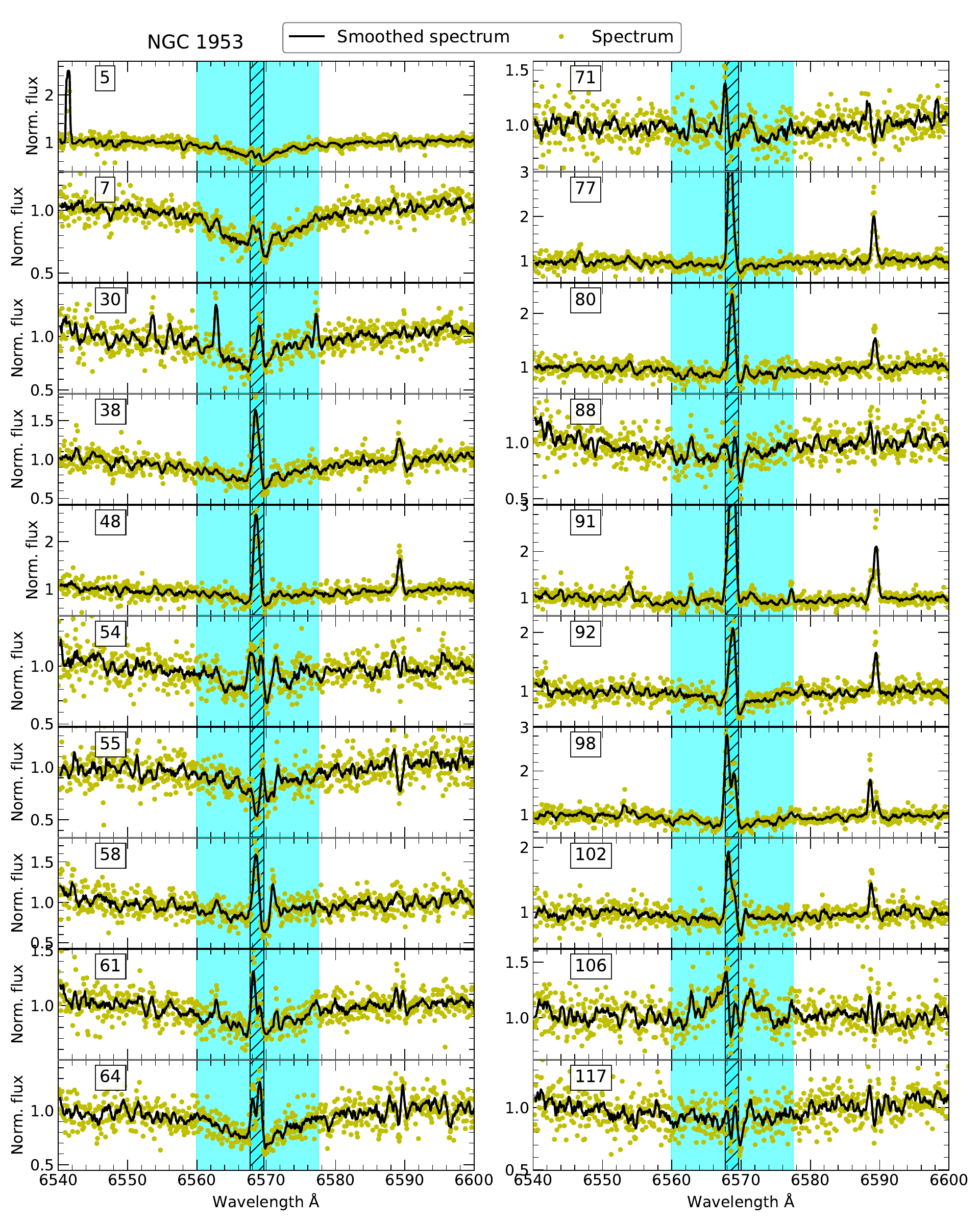}
	\caption{{\paul Same as Fig.~\ref{fig:ngc1856-spectra}} for NGC~1953.  The complete figure set (3 figures) is available in the online journal.}
	\label{fig:ngc1953-spectra}
\end{figure}

%

\figsetstart
\figsetnum{13}
\figsettitle{Sky correction on spetra recorded with M2FS for NGC~1856.}

\figsetgrpstart
\figsetgrpstart
\figsetgrpnum{13.1}
\figsetgrptitle{Target 3 to 33}
\figsetplot{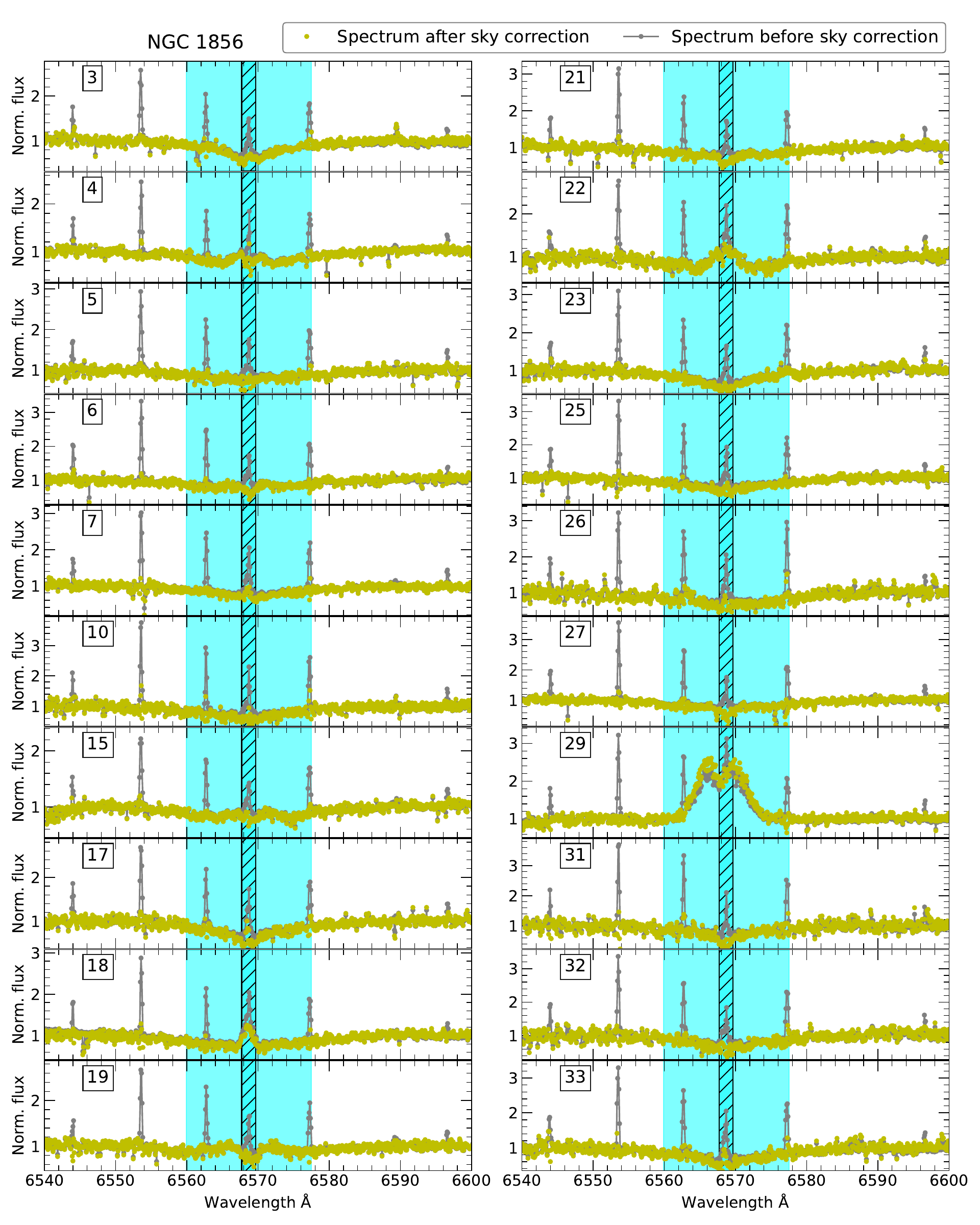}
\figsetgrpnote{}
\figsetgrpend

\figsetgrpstart
\figsetgrpstart
\figsetgrpnum{13.2}
\figsetgrptitle{Target 34 to 66}
\figsetplot{figures/ngc1856_plots_2_skycorr.pdf}
\figsetgrpnote{}
\figsetgrpend

\figsetgrpstart
\figsetgrpstart
\figsetgrpnum{13.2}
\figsetgrptitle{Target 67 to 102}
\figsetplot{figures/ngc1856_plots_3_skycorr.pdf}
\figsetgrpnote{}
\figsetgrpend

\figsetend

\begin{figure*}
	\centering
	\includegraphics[width=.99\linewidth]{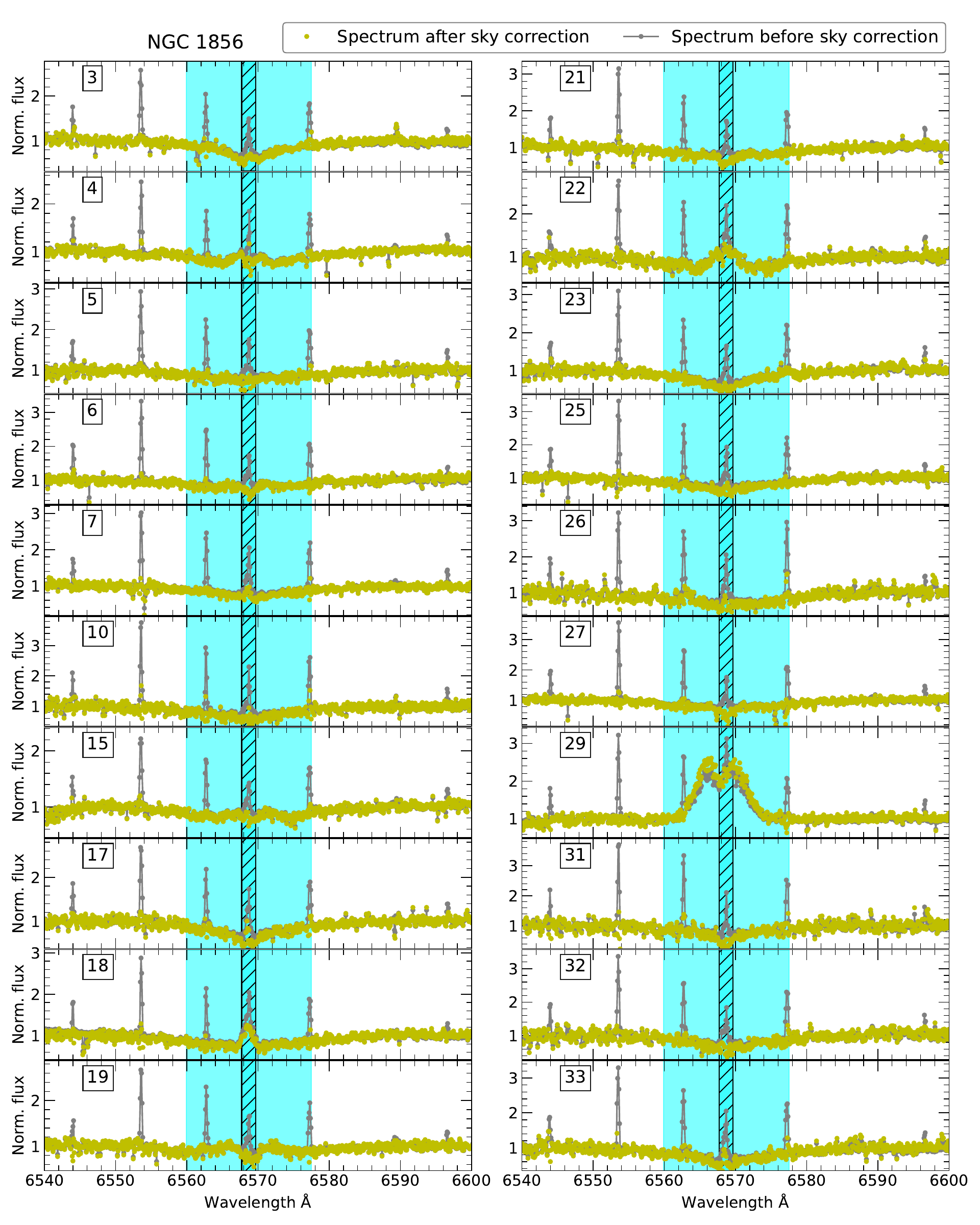}
	\caption{{\paul Sky correction on spectra recorded with M2FS for NGC~1856. The cyan regions indicates the width of the HST H$\alpha$ F656N filter centered on the observed H$\alpha$ line, as the radial velocity of the cluster causes a $\sim 260\kms$ shift between the observed H$\alpha$ emission and filter center. The hatched region indicates the typical width of the nebular emission recorded in the sky spectra. The gray and yellow points show the spectrum before and after sky correction. The complete figure set (3 figures) is available in the online journal.}}
	\label{fig:ngc1856-skycorr}
\end{figure*}

\bibliography{cfa}
\bibliographystyle{aasjournal}



\end{document}